\documentclass[a4paper,11pt]{article}
\pdfoutput=1 
\usepackage{jcappub} 
\usepackage[T1]{fontenc} 

\title{\boldmath Inflationary power spectra with quantum holonomy corrections}

\author{Jakub Mielczarek}

\affiliation{Institute of Physics, Jagiellonian University, Reymonta 4, 30-059 Cracow, Poland}
\affiliation{Department of Fundamental Research, National Centre for Nuclear Research,\\ 
Ho{\.z}a 69, 00-681 Warsaw, Poland}

\emailAdd{jakub.mielczarek@uj.edu.pl}

\abstract{In this paper we study slow-roll inflation with holonomy corrections  
from loop quantum cosmology.  Both tensor and scalar power spectra 
of primordial perturbations are computed up to the first order in slow-roll parameters
and $V/\rho_{c}$, where $V$ is a potential of the scalar field and $\rho_{c}$ 
is a critical energy density (expected to be of the order of the Planck energy density). 
Possible normalizations of modes at short scales are discussed. In case 
the normalization is performed with use of the Wronskian condition applied 
to adiabatic vacuum, the tensor and scalar spectral indices are not quantum 
corrected in the leading order. However, by choosing an alternative  
method of normalization one can obtain quantum corrections in the leading
order.  Furthermore, we show that the holonomy-corrected equation of motion 
for tensor modes can be derived from an effective background metric. 
This allows us to prove that the Wronskian normalization condition for the tensor 
modes preserves the classical form.}
\begin{document}
\maketitle
\flushbottom

\section{Introduction}

Effects of the quantum nature of space at the Planck scale predicted by loop quantum gravity 
(LQG) \cite{Ashtekar:2004eh} can be studied by introducing appropriate modifications at 
the level of the classical Hamiltonian. This so-called effective approach enables to relate some 
quantum gravitational phenomena with the realm of classical physics, which proved to be 
especially fruitful in the cosmological context, known as loop quantum cosmology (LQC) 
\cite{Bojowald:2006da,Ashtekar:2011ni}. 

As discussed in Ref. \cite{Bojowald:2012qya}, the effective approach in quantum gravity is 
conceptually similar to the effective approach in solid state physics. Namely, while calculations 
based on many-body Hamiltonian are extremely difficult to execute, there is a whole range of 
effective models enabling explanation of macroscopic phenomena in terms of atomic-scale 
physics. As an example, relevant for our further discussion, let us refer to the nature of refractive 
index $n$. In the effective model of frequency dependence of $n$, one considers a single 
atomic dipole interacting with electromagnetic plane wave. By virtue of homogeneity of a 
sample the formula for $n(\omega)$, characterizing macroscopic bulk, can be derived. The 
formula depends on known microscopic quantities as electron mass and elementary charge, 
but also contains some characteristic frequencies which cannot be derived from the model.  
The unknown values can be either fixed experimentally or derived from quantum mechanical
computations. 

In LQG, there exists an analogue of the many-body Hamiltonian in solid state physics. 
Action of this so-called Hamiltonian constraint on the spin network states, describing the state
gravitational field, is however not fully understood yet. When these difficulties are overcome
it will be possible to study some macroscopic or mesoscopic gravitational configurations 
numerically. This is in analogy to computations performed within condensed matter physics or
quantum chemistry. Meanwhile, the effective approach, competitive with the first-principle 
computations, can be utilized. 

Construction of effective models is facilitated by certain assumptions regarding 
symmetries of space. Here, we focus on homogeneous and isotropic background 
geometry\footnote{The same symmetries, but for crystal lattice, were applied in 
the mentioned effective model of refractive index.} described by the flat 
Freidmann-Robertson-Walker (FRW) metric on which inhomogeneities 
are considered perturbatively. Such setup is sufficient to study the generation of 
primordial perturbations during the phase of slow-roll inflation.    
 
Incorporation of LQG effects into cosmological models is performed by taking into account 
two types of corrections: inverse volume corrections and holonomy corrections. Both corrections 
reflect discrete nature of space at the Planck scale, however in a different manner.  While strength 
of inverse volume corrections depends on volume element, holonomy corrections are
sensitive to energy density. In case of inverse volume corrections, equations of motion for 
perturbations were derived in Ref. \cite{Bojowald:2008jv}. Based on this, corrections to the 
inflationary power spectra were derived in Ref. \cite{Bojowald:2010me}. The corrections were 
shown to be consistent with the 7-year WMAP data \cite{Bojowald:2011hd,Bojowald:2011iq}.

In this paper we focus on derivation of holonomy corrections to inflationary powers spectrum.  
As already mentioned, holonomy corrections are sensitive to energy density of matter. 
The characteristic energy scale, at which holonomy corrections are becoming important is 
\begin{equation}
\rho_c = \frac{3 m_{Pl}^2}{8\pi \gamma^2 L^2},
\end{equation}
where the Planck mass $m_{Pl} =1.22 \cdot 10^{19}$ GeV, $\gamma \sim \mathcal{O}(1)$ 
is the so-called Barbero-Immirzi parameter and $L$ is a length scale of the 
order of the Planck scale. Therefore, the critical energy density is expected 
to be of the order of the Planck energy density, $\rho_c \sim \rho_{Pl}$, where 
$ \rho_{Pl} \equiv m_{Pl}^4$. 

Holonomy corrections are modifying Friedmann equation into the following form 
\cite{Ashtekar:2006rx, Ashtekar:2006wn}
\begin{equation}
H^2 = \frac{8\pi}{3 m_{Pl}^2} \rho \left(1-\frac{\rho}{\rho_c} \right), 
\label{Friedmann}
\end{equation}
where $H$ is a Hubble factor and $\rho$ is energy density of the matter content. Positivity of the 
left hand side of this equation implies that energy density of matter is bounded from above, 
$\rho \leq \rho_c$. This leads to resolution of singularity problem of homogeneous cosmological 
modes. The big bang singularity is replaced by non-singular \emph{bounce}, which merges contracting 
and expanding phases \cite{Ashtekar:2006rx, Bojowald:2007zza}. 

Models of inflation are typically constructed with the use of scalar fields. In the simplest case 
it can be a single scalar field $\varphi$, the so-called inflaton field \cite{Linde:1983gd}. The 
single scalar is sufficient to construct a reliable model of the inflationary phase. Energy density 
of this field expresses as  
\begin{equation}
\rho = \frac{\dot{\varphi}^2}{2}+V(\varphi),
\end{equation}
where $V(\varphi)$ is a potential term. Equation of motion governing evolution of $\varphi$ is not a 
subject of holonomy corrections and takes the standard form:
\begin{equation}
\ddot{\varphi}+3H\dot{\varphi}+V_{,\varphi}=0.
\label{KGeq}
\end{equation}

The holonomy corrections can be also introduced into equations governing evolution of 
cosmological perturbations. In particular, it was found that, while holonomy corrections 
are present, equation of motion for the Mukhanov variable $v$ is\cite{Cailleteau:2011kr}:
\begin{equation}
\frac{d^2}{d\tau^2} v-\Omega \nabla^2 v - \frac{z^{''}_S}{z_S} v = 0.  
\label{MukhanovEQ}
\end{equation}
Here $\tau$ is a conformal time defined as $d\tau = dt /a$. Moreover, the 
holonomy correction function  
\begin{equation}
\Omega = 1-2\frac{\rho}{\rho_c} 
\label{OmegaDef}
\end{equation}
and 
\begin{equation}
z_S = a  \frac{\dot{\varphi}}{H}. 
\end{equation}
It is clear that, while $\rho \ll \rho_{c}$ the classical expression with $\Omega=1$  is correctly recovered. 
 
Based on the Mukhanov variable $v$, perturbations of curvature $\mathcal{R} = \frac{v}{z}$ 
can be derived. The quantity $\mathcal{R}$ is a key object characterizing scalar perturbations,
allowing for computation of the scalar power spectrum.

The equation (\ref{MukhanovEQ}) was originally derived by considering requirements of 
anomaly freedom for the scalar perturbations \cite{Cailleteau:2011kr}. Later, it was shown
that this equation can also be obtained from the lattice loop quantum cosmology 
\cite{WilsonEwing:2011es,WilsonEwing:2012bx}. The equation (\ref{MukhanovEQ})
can be seen as a result of discretization of space for homogeneous cubic cells with the 
lattice spacing $L$. Cosmological consequences of equation (\ref{MukhanovEQ}) have 
not been studied in details yet. As an interesting application, power spectra from a matter 
bounce was found \cite{WilsonEwing:2012pu}.

For tensor modes (gravitational waves), holonomy corrected version of the equation is 
\cite{Cailleteau:2012fy} 
\begin{equation}
\frac{d^2}{d\tau^2}h_{i} +
2\left( \mathcal{H}-\frac{1}{2\Omega} \frac{d\Omega}{d\tau} \right)\frac{d}{d\tau}h_{i}
-\Omega \nabla^2h_{i}= 0, 
\label{TensorEOM1}
\end{equation}
where $i=\otimes, \oplus$ corresponds to two polarizations of gravitational waves. 
This equation can be rewritten into the form 
\begin{equation}
\frac{d}{d\tau^2}u-\Omega \Delta u-\frac{z_T ^{''}}{z_T}u=0,
\label{TensorEOM2}
\end{equation}
where $z_T = a/\sqrt{\Omega}$ and $u =\frac{a h_{\otimes, \oplus}}{\sqrt{16 \pi G} \sqrt{\Omega}}$. 
The equation (\ref{TensorEOM1}) differs from the equation of motion for tensor modes with 
holonomy corrections originally derived in Ref. \cite{Bojowald:2007cd}. This is because the original  
derivation has not been based on anomaly freedom Hamiltonian.  However, taking into account the 
issue of anomaly freedom became possible thanks to analysis of scalar perturbations with holonomy 
corrections performed in Ref. \cite{Cailleteau:2011kr}.    

So far, the equation (\ref{TensorEOM1}) was applied to study generation of tensor perturbation 
across the cosmic bounce \cite{WilsonEwing:2012pu, Linsefors:2012et}. Nevertheless, there 
is a whole aggregation of previous analyses performed with the use of the original equation for 
tensor modes with holonomy corrections (See e.g. \cite{Copeland:2008kz, Grain:2009kw, 
Mielczarek:2010bh, Mielczarek:2008pf}).  There were also earlier attempts to study holonomy 
corrections for scalar perturbations. However, they were not consistent with the requirement of 
anomaly freedom. In particular, the studies for scalar perturbations were performed in 
Ref. \cite{Artymowski:2008sc}. There is also an alternative approach to incorporate 
loop quantum corrections to cosmological perturbations developed in Refs. \cite{Agullo:2012sh,
Agullo:2012fc,Agullo:2013ai}.  

It is worth mentioning at this point that because we consider model with the scalar matter, 
the vector modes are not activated and identically equal zero \cite{Mielczarek:2011ph}.

For both tensor and scalar perturbations the deformation factor $\Omega$ is placed in 
front of Laplace operator. Therefore, it can be considered as an effective speed of light 
squared. Namely, by neglecting the cosmological factor and assuming the plane wave 
solution $v \propto e^{i( {\bf k \cdot x} -\omega \tau)}$, we find the following dispersion 
relation $\omega^2 = \Omega k^2$, based on which the phase velocity $v_{ph} = 
\frac{\omega}{k}= \sqrt{\Omega}$.  Therefore, the refractive index 
\begin{equation}
n \equiv \frac{1}{v_{ph}} = \frac{1}{\sqrt{\Omega}}. 
\end{equation}
While $\rho \rightarrow \rho_c/2$ ($\Omega \rightarrow 0$) the refractive index becomes 
infinite, and speed of propagation tends to zero. As discussed in Ref. \cite{Mielczarek:2012tn} 
this can be associated with the state of \emph{asymptotic silence}.  At the energy densities 
$\rho \in ( \rho_c/2, \rho_c]$ the refractive index is purely imaginary. As discussed 
in Ref. \cite{Linsefors:2012et} this not necessarily means that space 
is opaque for the propagation of waves. The waves are not only evanescent in this region, 
but can be amplified as well. Behavior observed from the numerical computations differs 
with the intuition gained from \emph{e.g} analysis of  waves in plasma with 
frequencies lower than the plasma frequency.  Furthermore, as discussed in 
Refs. \cite{Bojowald:2011aa,Mielczarek:2012pf} the region of negative $\Omega$ can be 
associated with the change of metric signature from Lorentzian to Euclidean one.    
  
However, in our calculations of inflationary power spectra, we restrict ourselves to the 
regime where $\Omega>0$. Therefore, the interesting behavior in vicinity of $\Omega=0$ 
and at the negative values of $\Omega$ will not be relevant.  We will come back to
the issue of evolution of modes in the  $\Omega \leq 0$ in our further studies.  
     
\section{Slow-roll inflation}

During the slow-roll roll inflation Universe underwent an almost exponential expansion.  
The deviation from the exponential (de Sitter) growth of the scale factor is parametrized 
by the slow-roll parameters, which are much smaller than unity. The slow-roll roll 
inflation is characterized by gradual decreasing of $\varphi$ in a potential $V(\varphi)$. 
In this regime, energy density of the scalar field is dominated by its potential energy, 
therefore $\dot{\varphi}^2 \ll V(\varphi)$. Because of that, the modified Friedmann equation 
(\ref{Friedmann}) can be approximated by 
\begin{equation}
H^2 \simeq \frac{8\pi}{3 m_{Pl}^2} V \left(1-\frac{V}{\rho_c} \right). 
\label{FriedmannSR}
\end{equation}
Furthermore, flatness of the potential implies that $\ddot{\varphi}$ in equation (\ref{KGeq}) can be 
neglected, such that 
\begin{equation}
3H\dot{\varphi}+V_{,\varphi}\simeq 0.
\label{KGSR}
\end{equation}
Using (\ref{KGSR}) to eliminate $\dot{\varphi}$ from the condition $\dot{\varphi}^2 \ll V(\varphi)$
and by using  (\ref{FriedmannSR}), one can define  \cite{Artymowski:2008sc}
\begin{equation}
\epsilon :=  \frac{m_{Pl}^2}{16 \pi} \left( \frac{V_{,\varphi}}{V} \right)^2 \frac{1}{(1-V/\rho_c)} 
= - \frac{\dot{H}}{H^2} \frac{1}{(1-V/\rho_c)},
\label{SRepsilon}
\end{equation}
such that $\epsilon \ll 1$ for the slow-roll inflation. 

By differentiating the slow-roll equation $\dot{\varphi} \simeq - \frac{V_{,\varphi}}{3H}$ we find
\begin{equation}
\ddot{\varphi} = - \frac{V_{,\varphi\varphi} \dot{\varphi} }{3H}+\frac{V_{,\varphi} \dot{H} }{3H^2}.
\end{equation} 
Because $|\ddot{\varphi}| \ll | V_{,\varphi} |$, the absolute value of 
\begin{equation}
\frac{\ddot{\varphi}}{V_{,\varphi}} \simeq  \frac{1}{3}\eta - \frac{1}{3}  \epsilon \left( 1- \frac{V}{\rho_c} \right)
\label{eqvpvp}
\end{equation}
has to be much smaller than unity. Following Ref. \cite{Artymowski:2008sc} let us introduce 
the second slow-roll parameter
\begin{equation}
\eta :=  \frac{m_{Pl}^2}{8 \pi} \left( \frac{V_{,\varphi\varphi}}{V} \right)  \frac{1}{(1-V/\rho_c)},
\end{equation}
satisfying $|\eta|  \ll 1$ for $|\ddot{\varphi}| \ll | V_{,\varphi} |$. Based on (\ref{eqvpvp}) we can also define 
\begin{equation}
\delta := \eta - \epsilon \left( 1- \frac{V}{\rho_c} \right) = - \frac{\ddot{\varphi}}{H\dot{\varphi}},
\label{deltaSR}
\end{equation}
satisfying $\delta \ll 1$. 

While studying cosmological perturbations it is convenient to work with the 
conformal time $\tau \equiv \int \frac{dt}{a}$. Here, it is defined such that $\tau \in (-\infty,0)$.
Based on the definition of conformal time and integrating by parts, we find 
\begin{equation}
\tau = \int \frac{dt}{a} = \int \frac{da}{a^2 H} = -\frac{1}{Ha}-\int \frac{1}{a} \frac{\dot{H}}{H^2} dt
= -\frac{1}{Ha}+\tau \epsilon \left(1-\frac{V}{\rho_c} \right),
\end{equation}
where in the last equality we applied (\ref{SRepsilon}). This enables us to write expression for 
the time dependence of the scale factor 
\begin{equation}
a = - \frac{1}{H \tau} \frac{1}{\left[1- \epsilon\left(1- \frac{V}{\rho_c}\right)\right]}.
\end{equation}

In the slow-roll regime the $\Omega$ function, defined in Eq. \ref{OmegaDef}, is 
approximated by 
\begin{equation}
\Omega \simeq 1 - 2 \delta_H,
\end{equation}
where for the later convenience we introduced parameter
\begin{equation}
\delta_H := \frac{V}{\rho_c}. 
\end{equation}
This parameter reflects deviation from the classical slow-roll inflation due to 
holonomy corrections. In the classical limit, which corresponds to 
$\rho_c \rightarrow \infty$, $\delta_H$ goes to zero. In what follows 
we will consider only linear corrections in $\delta_H$. This is because 
$\delta_H$ is expected to be a very small quantity. The assumption that 
the slow-roll regime takes place in the Lorentzian regime ($\Omega > 0$)
implies that $\delta_H < 1/2$. One can however motivate that $\delta_H \ll 1/2$ 
unless the critical energy density $\rho_c$ is not much smaller than 
the Planck energy density. As an example, let us consider model with a
massive potential $V(\varphi)=\frac{1}{2} m^2 \varphi^2$. Taking the 
inflaton mass $m \sim 10^{-6}\ m_{Pl}$ and value of the scalar field 
$\varphi \sim m_{Pl}$ in agreement with cosmological observations 
one can estimate that $V(\varphi) \sim 10^{-12} \rho_{Pl}$. Therefore, 
for $\rho_c \sim \rho_{Pl}$ one can expect that $\delta_H$ has the extremely 
small value $\delta_H \sim 10^{-12}$.   On the other hand if $\rho_c \sim 10^{-12} 
\rho_{Pl}$ or smaller, the holonomy corrections are becoming observationally 
relevant and allow to constraint models with low critical energy density.  

Here, we keep terms linear in both slow-roll parameters and $\delta_H$ as 
well as the mixed terms $\mathcal{O}(\epsilon \delta_H)$ and $\mathcal{O}(\eta \delta_H)$. 
Contribution from the second order expansion in the slow-roll parameters 
is not taken into account.  However, in case the $\delta_H$ is extremely 
small, as estimated above, the terms $\mathcal{O}(\epsilon^2)$ will 
dominate contributions from $\mathcal{O}(\epsilon \delta_H)$. The 
derived expressions will therefore have practical application only 
to the regime where $\rho_{c} \in (\sim 0.01, \sim 1/2)$\footnote{Under 
assumption that the potential of the inflaton field is quadratic.}, 
where the lower limit comes from estimating values of the slow-roll parameters. 
The estimated range overlaps with the domain which can be probed with use of 
currently available observational data. Furthermore, the theoretical predictions 
performed here will set a stage for more comprehensive considerations 
of the second order expansion in the slow-roll parameters. This will 
extend a range of testable values of $\rho_c$, of course if it is allowed by 
observational data.

\section{Normalization of modes}

In this section we will present some possible choices of the short scale 
normalizations for the perturbations with holonomy corrections. In must 
be stressed that we do not explore any representative class of states. 
However, the considered normalizations seem to be the most reliable and 
best physically motivated. Because of this ambiguity, the choice of the 
normalization is the weakest point of a whole construction of the model 
of generation of primordial perturbations during the inflationary phase. 
This concerns also the case without quantum gravitational corrections. 
Therefore, here we pay a lot of attention to this issue. 

Performing the Fourier transform $v({\bf x},\tau) = \int \frac{d^3 x}{(2\pi)^{3/2}}  v_k(\tau) e^{i{\bf k \cdot x}} $ 
of the equations of modes for tensor and scalar perturbations we find  
\begin{equation}
\frac{d}{d\tau^2}v_k+\Omega k^2 v_k-\frac{z^{''}}{z}v_k=0, 
\label{modeeq}
\end{equation}
where $k^2 = {\bf k \cdot k}$, and expression on $z$ depends on whether scalar or tensor 
mods are studied.  In both cases  $\frac{z^{''}}{z} \approx \mathcal{H}^2 \simeq \frac{1}{\tau^2}$. 
Based on this, one can define super-horizontal limit when $ \sqrt{\Omega}k \ll \mathcal{H}$ 
and short scale limit when  $\sqrt{\Omega} k  \gg \mathcal{H}$. In the super-horizontal 
limit the $\Omega k^2 v_k$ factor in Eq. \ref{modeeq} can be neglected and an approximate 
solution $v_k = c_1 z + c_2\int^{\tau}\frac{d\tau'}{z}$ can be found. Because the physical 
amplitudes of perturbations are proportional to the ratio $v_k/z$, it is clear that amplitudes
are ``frozen'' at the super-horizontal scales. This process beings when  $ \sqrt{\Omega}k \approx 
\mathcal{H}$\footnote{It is worth noticing that this condition differs from the classical one $k \approx 
\mathcal{H}$ due to presence of time dependent function $\Omega$. Furthermore, it is worth 
to stress that $\frac{z^{''}}{z}$ is a $\Omega$-dependent function leading corrections
in expression $\frac{z^{''}}{z} \approx \mathcal{H}^2 \simeq \frac{1}{\tau^2}$. In particular, for the 
tensor modes $\frac{z_T^{''}}{z_T} = \mathcal{H}^2\left[2-\epsilon(1+5\delta_H)+\dots \right]$. 
The correction due to $\delta_H$ contributes however together with the $\epsilon$ factor, contrary 
to the contribution $\Omega=1-2\delta_H+\dots$ in front of the $k^2$ factor.}. In the short scale limit, 
the factor $\frac{z^{''}}{z}u_k$ in Eq. \ref{modeeq} can be neglected and the equation for modes 
reduces to
\begin{equation}
\frac{d^2}{d\tau^2} v_k+\Omega(\tau) k^2 v_k \approx 0.
\label{UVEOMdef}
\end{equation}
For the slow-roll inflation, the $\Omega$ is only weakly dependent on $\tau$ and solution to 
equation (\ref{UVEOMdef}) can be found by applying the WKB approximation. We find that 
\begin{equation}
v_k = \frac{c_1}{\sqrt{2k\sqrt{\Omega}} }e^{-ik\int^{\tau}\sqrt{\Omega(\tau')}d\tau'}+
\frac{c_2}{\sqrt{2k\sqrt{\Omega}}}e^{+ik\int^{\tau}\sqrt{\Omega(\tau')}d\tau'},
\label{WKBHolo}
\end{equation}
which is superposition of plane waves traveling forward ($e^{-i k \int^{\tau}\sqrt{\Omega(\tau')}d\tau'}$) and 
backward ($e^{+i k \int^{\tau}\sqrt{\Omega(\tau')}d\tau'}$)  in time.  Validity of the WKB approximation requires that 
\begin{equation}
\frac{1}{2}\left| \frac{\Omega'}{\Omega} \right|  \frac{1}{k\sqrt{\Omega}} \ll 1. 
\end{equation}
Because we are in the short scale limit $\sqrt{\Omega}k \gg \frac{1}{|\tau|}$ and 
\begin{equation}
\frac{\Omega^{'}}{\Omega} = - \frac{4\epsilon }{\tau} \delta_H,
\label{OmegaPrimeOmega}
\end{equation}
the condition of validity of the WKB approximation simplifies to $\epsilon \delta_H < 0$. Because 
both $\epsilon$ and $\delta_H$ are smaller than unity for the considered slow-roll evolution, 
the WKB approximation (\ref{WKBHolo}) holds. 

Canonical commutation relation between quantum field $\hat{v}$ and its conjugated 
momenta requires Wronskian condition
\begin{equation}
W(v_k,v_k') \equiv v_k \frac{d v^{*}_k}{d\tau}-v^{*}_k \frac{d v_k}{d\tau} = i
\label{Wronskian}
\end{equation}
to be satisfied. This is the usual way the modes are normalized and also the 
place where quantum mechanics enters into description of primordial 
perturbations. The Wronskian condition applied to solution (\ref{WKBHolo})
leads to relation
\begin{equation}
|c_1|^2-|c_2|^2=1.
\end{equation}
The initial four numbers ($c_1, c_2 \in \mathbb{C}$) parametrizing solution (\ref{WKBHolo}) 
are therefore reduced to three. Because the total phase is physically irrelevant, the family of 
normalized solutions in the short scale limit is characterized by two real numbers. Their 
values  have to be fixed by hand. The obtained solutions are used to normalize 
general solutions to the equations of motion.  

It is worth stressing at this point that while considering the short scale limit $ \sqrt{\Omega}k \ll \mathcal{H}$ 
one has to be cautious about the limit $k\rightarrow \infty$. Such limit can be performed only 
formally because for $k > \frac{a}{l_{Pl}}$ the classical description of space is expected to be 
no more valid due to quantum gravitational effects.  We do not consider such \emph{trans-Planckian modes} here. 

\subsection{$\Omega-$deformed Minkowski vacuum} 

For the particular choice $c_1=1$, solution (\ref{WKBHolo}) contains incoming modes only:
\begin{equation}
v_k = \frac{1}{\sqrt{2k\sqrt{\Omega}}}e^{-ik\int^{\tau}\sqrt{\Omega(\tau')}d\tau'}.
\label{WKBHoloMink}
\end{equation}
This solution reduces to the so-called Minkowski (Bunch-Davies) vacuum
\begin{equation}
v^{M}_k \equiv \frac{e^{-i k \tau }}{\sqrt{2k}} 
\end{equation}
in the classical limit ($\Omega \rightarrow 1$),  which has been extensively 
used to normalize cosmological perturbations.

Based on (\ref{WKBHoloMink}) we define $\Omega-$deformed Minkowski vacuum 
to be 
\begin{equation}
v^{\Omega M}_k \equiv \frac{e^{-ik\int^{\tau}\sqrt{\Omega(\tau')}d\tau'}}{\sqrt{2\sqrt{\Omega}k}}
\approx v^{M}_k\left[1+ \left(\frac{1}{2}+ik\tau \right) \delta_H +\mathcal{O}(\delta_H^2)\right],  
\label{OmegaMdefExp}   
\end{equation}
where in the second equality we neglected time variation of $\Omega$.  

\subsection{Bojowald-Calcagni normalization}

Another possibility of normalizing modes was proposed in Ref. \cite{Bojowald:2010me} 
for the case of perturbations with inverse volume corrections. The proposal made by Bojowald 
and Calcagni was that in the short scale limit the solution to (\ref{UVEOMdef}) can be written 
up to the first order in $\delta_H$ as follows 
\begin{equation}
v_{k}^{BC}= v^M_k(1+y(k,\tau) \delta_H), 
\label{vBC1}
\end{equation}
where $y(k,\tau)$ is some unknown function\footnote{In the original paper \cite{Bojowald:2010me} 
the expansion was performed not in therms of $\delta_H$ but $\delta_{Pl}$ relevant for inverse volume 
corrections.}. By plugin in (\ref{vBC1}) to (\ref{UVEOMdef}) we find the following equation for the function $y$:
\begin{equation}
y^{''}-2(2\mathcal{H}\epsilon+i k )y'+(4i \mathcal{H}\epsilon k-2\epsilon \mathcal{H}^2-2\delta_H k^2 )y-2k^2 = 0,
\label{BCEOMy1}
\end{equation}
where we used relations
\begin{eqnarray}
\delta_H' &=&  -2\epsilon  \mathcal{H} \delta_H, \\
\delta_H^{''} &=& -2\epsilon  \mathcal{H}^2 \delta_H.
\end{eqnarray}
The equation (\ref{BCEOMy1}) requires certain simplifications. Firstly, because we are interested in the 
first order correction in $ \delta_H$ we can skip the factor $-2\delta_H k^2y$  in  (\ref{vBC1}),  which 
would generate higher order contribution. Secondly, because we are looking for the short scale solution 
($\sqrt{\Omega}k \gg \mathcal{H}$) the factor $-2\epsilon \mathcal{H}^2y$ can be ignored as well. 
This second approximation turns out to be beneficial while searching for analytic solution to the equation of 
motion for $y$. The reduced equation (\ref{BCEOMy1}) is now 
\begin{equation}
y^{''}-2(2\mathcal{H}\epsilon+i k )y'+4i \mathcal{H}\epsilon k y -2k^2 = 0.
\label{BCEOMy2}
\end{equation}
For $\epsilon = 0$, solution to this equation is
\begin{equation}
y = i k \tau +c_1+c_2 e^{2 i k\tau}, 
\end{equation}
where $c_1$ and $c_2$ are constants of integration. Because the $e^{2 i k\tau}$ factor 
would lead to outgoing modes we fix $c_2=0$. Value of the factor $c_1$ can be determined 
by considering the case $\epsilon \neq 0$. Let us now find solution in the form $y=a+bx$. 
By applying it to (\ref{BCEOMy2}), where $\mathcal{H} = - \frac{1}{\tau}$\footnote{Here we 
use the simplified de Sitter solution instead of $\mathcal{H} = - \frac{1}{\tau}\cdot 
\frac{1}{\left[1-\epsilon(1-\delta_H)]\right]}$, which is sufficient within the considered 
order of approximation.}, we find special solution $y=a+bx$ with $a=\frac{1}{1+2\epsilon}$ 
and $b=\frac{i}{1+2\epsilon}$ to equation (\ref{BCEOMy2}).  Requirement of analytic continuity 
of the solution (in respect to $\epsilon$) between the cases $\epsilon=0$ and $\epsilon\neq0$ fixes 
the value of $c_1$. In consequence, we obtain  
\begin{equation}
v^{BC}_k = v^{M}_k\left[1+ \frac{1}{1+2\epsilon} \left(1+ik\tau \right) \delta_H +\mathcal{O}(\delta_H^2)\right]  
\simeq  v^{M}_k\left[1+\left(1+ik\tau \right) \delta_H\right]. 
\end{equation}
It is worth noticing a slight difference between this case and predictions of the $\Omega-$deformed Minkowski
vacuum (\ref{OmegaMdefExp}).  In contrast to that case, the method presented in this subsection does not utilize 
the Wronskian condition in order to normalize the mode functions.  This may have advantages if we have reason 
to suppose that the Wronskian condition is deformed but the form of deformation is not known.  

\subsection{Deformed Wronskian condition}

Let us suppose that indeed the Wronskian condition is deformed due to presence of $\Omega$. 
Such deformation can come from the fact that, because $\Omega$ is present in equations of motion, 
inner product must differ from the classical one. We will discuss this issue in more details in the 
next section, while here wa assume the classical Wronskian condition (\ref{Wronskian}) is 
deformed to 
\begin{equation}
W_{\Omega}(v_k,v_k')= v_k \frac{d v^{*}_k}{d\eta}-v^{*}_k \frac{d v_k}{d\eta} = i f(\Omega),
\end{equation}
where $f(\Omega)$ is some function of $\Omega$, defined such that $\lim_{\Omega  \rightarrow1} =1$. 
In case we have no hints what the functional form of $f(\Omega)$ is we can investigate a 
power-low parametrization  
\begin{equation}
f(\Omega) = \Omega^n. 
\end{equation}
In this case, the counterpart of (\ref{OmegaMdefExp}) is 
\begin{equation}
v_{k}^{(n)} \equiv \frac{e^{-i \sqrt{\Omega}k \tau }}{\sqrt{2k \Omega^{1/2-n}}}
=v_{M}\left[1+ \left(\frac{1}{2}-n+ik\tau \right) \delta_H +\mathcal{O}(\delta_H^2)\right].     
\end{equation}
As we see, for $n=-\frac{1}{2}$, the normalization from the deformed Wronskian 
condition overlaps with Bojowald-Calcagni normalization up to the first order in $\delta_H$:
\begin{equation}
v_{k}^{(-1/2)} \simeq v^{BC}_k.  
\end{equation}

\section{Inner product and the Wronskian condition}

Let us now address the issue of validity of the Wronskian normalization in presence 
of holonomy corrections in more details. 

For the a pair of fields $\phi_1, \phi_2$ satisfying Klein-Gordon equation $(\Box-m^2)\phi =0$, the inner product is 
\cite{Wald1994}  
\begin{equation}
\langle \phi_1 | \phi_2 \rangle := i \int_{\Sigma} d^3x \sqrt{q} n^{\mu} 
\left( \phi_2^* \partial_{\mu} \phi_1 - \phi_1 \partial_{\mu} \phi_2^* \right),
\label{InnerProduct}
\end{equation}
where $n^{\mu} $  is  a future-direction unit ($g_{\mu\nu} n^{\mu} n^{\mu} =-1$) vector normal to Cauchy 
surface $\Sigma$  and $q$ is a determinant of the spatial metric on $\Sigma$. To remind, the Cauchy surface 
is a spatial hypersurface at which the initial conditions are imposed.  The inner product (\ref{InnerProduct}) is 
defined such that  it does not depend on the choice of a Cauchy surface:
\begin{equation}
\langle \phi_1 | \phi_2 \rangle_{\Sigma_1} =  \langle \phi_1 | \phi_2 \rangle_{\Sigma_2}. 
\label{IPCauchyInd}
\end{equation}
The proof is direct and employs a Gauss law, Klein-Gordon equation and vanishing of
$\phi $ at spatial infinities (See e.g. Ref. \cite{Ford:1997hb}). It is also worth noticing, that the 
inner product (\ref{InnerProduct}) is not positive-definite. 
  
In the case studied in this paper, the Klein-Gordon equations for  scalar ad tensor perturbations are deformed 
with respect to the classical one. Therefore, in general, one could expect that (\ref{InnerProduct}) is not a good scalar
product because the condition (\ref{IPCauchyInd}), requiring the  Klein-Gordon equation to be satisfied, is  
not fulfilled. This can imply that the Wronskian condition (\ref{Wronskian}), resulting from normalization 
of modes with use of  (\ref{InnerProduct}), is deformed. 

However, if we manage to find an effective metric $ g^{eff}_{\mu\nu}$ which leads to holonomy deformations
of the equations of perturbations, then the proof the condition (\ref{IPCauchyInd}) would remain in force, 
and the inner product (\ref{InnerProduct}) can be used. In what follows we show that such construction is 
possible for tensor perturbations. 
  
For any component $\phi$ of the tensor perturbations, the equation of motion is 
\begin{equation}
\frac{d^2}{d\tau^2}\phi+2\left( \mathcal{H}-\frac{1}{2\Omega} \frac{d\Omega}{d\tau} \right)\frac{d}{d\tau}\phi
-\Omega \nabla^2\phi= 0.
\label{KGefftau}
\end{equation}
In the coordinate time ($dt=a d\tau$) this equation can be written as 
\begin{equation}
\ddot{\phi}+3H \dot{\phi}- \frac{\dot{\Omega}}{\Omega} \dot{\phi}- \frac{\Omega}{a^2}\nabla^2 \phi=0. 
\label{KGefft}
\end{equation}
The classical Klein-Gordon equation  $\Box\phi =0$ (the tensor modes are massless) on the FRW 
background is recovered by taking $\Omega=1$.  It can be proved by direct calculation, that the 
holonomy corrected equations for tensor modes can be derived from the wave equation $\Box\phi =0$
at the effective FRW background given by the line element 
\begin{equation}
ds^2_{eff} = g^{eff}_{\mu\nu}dx^{\mu} dx^{\nu} =  -\sqrt{\Omega} N^2 dt^2
+\frac{a^2}{\sqrt{\Omega}} \delta_{ab}da^adx^b,  \label{effectivmetric}
\end{equation}
where $N$ is a lapse function. In particular for the coordinate time ($N=1$) we have  
\begin{eqnarray}
\Box \phi &=& \nabla^{\mu} \nabla_{\mu} \phi = \frac{1}{\sqrt{-g}} \partial_{\mu}(\sqrt{-g} g^{\mu\nu}  \partial_{\nu} \phi  ) \nonumber \\
&=& - \frac{1}{\sqrt{\Omega}} \ddot{\phi}+ \frac{\sqrt{\Omega}}{a^2}\nabla^2 \phi 
- \frac{1}{\sqrt{\Omega}}\left(3H- \frac{\dot{\Omega}}{\Omega} \right)\dot{\phi},     
\label{Boxeff}
\end{eqnarray}
where we used $g_{\mu\nu} =  g^{eff}_{\mu\nu}$.  By equating (\ref{Boxeff}) to zero and 
multiplying by $-\sqrt{\Omega}$, the equation (\ref{KGefft}) is recovered. 

It is worth noticing that the effective metric (\ref{effectivmetric}) is conceptually similar to 
\emph{dressed metric} approach to quantum fields on quantum spaces \cite{Ashtekar:2009mb,Agullo:2012sh}. 
In our case, quantum gravitational effects are ``dressing'' the FRW metric leading to the 
effective metric (affected by $\Omega$ terms), which is felt by test fields. 

Now we can check if the Wronskian condition derived based on (\ref{InnerProduct}) holds the classical 
form.  We will be interested in the form of the Wronskian condition for the field
\begin{equation}
u = \frac{a}{\sqrt{\Omega}} \phi, 
\end{equation}
which, as can be seen by substituting to (\ref{KGefftau}), fulfills equation 
\begin{equation}
\frac{d^2}{d\tau^2}u-\Omega \nabla^2 u- \frac{z^{''}}{z}u=0, 
\end{equation}
where $z=a/\sqrt{\Omega}$.  Because the Cauchy hypersurface is spatial, based on $g_{\mu\nu} n^{\mu} n^{\mu} =-1$, 
we find $n^0 = \frac{1}{N \Omega^{1/4}}$ and $n^a=0$, which gives us $n^{\mu}\partial_{\mu}=\frac{1}{N \Omega^{1/4}} \partial_{t}$.
Then, for the conformal time ($N=a$), the inner product of two fileds $\phi$ is 
\begin{eqnarray}
\langle \phi | \phi \rangle &=& i \int_{\Sigma} d^3x \sqrt{q} \left( \phi^* n^{\mu}\partial_{\mu} \phi - \phi n^{\mu}\partial_{\mu} \phi^* \right)
\nonumber \\
&=&  i \int_{\Sigma} d^3x   \frac{a^3}{\Omega^{3/4}} \frac{1}{a \Omega^{1/4}} \frac{\Omega}{a^2} 
\left( u^*  \frac{du}{d\tau} - u\frac{du^*}{d\tau}  \right) \nonumber  \\
&=&  i \int_{\Sigma} d^3x \left( u^*  \frac{du}{d\tau} - u\frac{du^*}{d\tau}  \right) \nonumber  \\
&=&-i \int_{\Sigma} d^3x W(u,u') =  \int_{\Sigma} d^3x = V_0 =1,
\end{eqnarray}
where the classical Wronskian condition (\ref{Wronskian}) was used to get the proper normalization $\langle \phi | \phi \rangle =1$.
Here, we assumed that the spatial integration is restricted to $V_0$, or equivalently the spatial topology is compact and 
has coordinate volume $V_0$. This volume can be conventionally fixed to one.  Alternatively the field $\phi$ can be rescaled by
$\phi \rightarrow  \frac{1}{\sqrt{V_0}} \phi $ to compensate the contribution from the spatial integration over $V_0$. 

In summary, for the tensor modes, the inner product (\ref{InnerProduct}) is properly defined and normalization condition
$\langle \phi | \phi \rangle =1$ leads to the  classical Wronskian condition (\ref{Wronskian}).  The $\Omega-$deformed 
vacuum seems  to therefore be the right choice for the tensor modes.  It remains to show if the similar construction can 
be performed for the scalar modes as well. 
 
\section{Tensor power spectrum}

In this section we will compute inflationary tensor power spectrum with holonomy corrections. 
Starting point for our considerations is the equation  
\begin{equation}
\frac{d^2}{d\tau^2}u_k+\Omega k^2 u_k- \frac{z_T ^{''}}{z_T}u_k=0, 
\end{equation}
where $k^2 = {\bf k \cdot k}$. Having solutions for $u_T$ and $z_T$, tensor power spectrum 
can be found from the definition 
\begin{equation}
\mathcal{P}_T(k) = 64 \pi G \frac{k^3}{2\pi^2} \left| \frac{u_k}{z_T} \right|^2.
\label{TensorPower}
\end{equation}

With use of $z_T = a/\sqrt{\Omega}$, expression for the effective mass term can be written as   
\begin{equation}
m^2_{eff} \equiv -\frac{z_T ^{''}}{z_T}=- \frac{a^{''}}{a}+ \frac{a^{'}}{a}\frac{\Omega^{'}}{\Omega}
+\frac{1}{2}\frac{\Omega^{''}}{\Omega} -\frac{3}{4}\left(\frac{\Omega^{'}}{\Omega} \right)^2.
\end{equation}
All the factors contributing to $m^2_{eff}$ can be expressed in terms of conformal time $\tau$ 
as well as $\epsilon, \eta$ and $\delta_H$. With use of the slow-roll conditions, these terms are:
\begin{eqnarray}
\frac{a^{'}}{a} &=& - \frac{1}{\tau} \left[1+\epsilon\left(1- \delta_H\right)\right],  \\
\frac{a^{''}}{a} &=& \frac{1}{\tau^2}\left[2+3\epsilon\left(1- \delta_H\right)\right], \\
\frac{\Omega^{''}}{\Omega} &=& \frac{4\epsilon }{\tau^2} \delta_H. 
\end{eqnarray}
The expression for $\frac{\Omega^{'}}{\Omega}$ is given in Eq. \ref{OmegaPrimeOmega}. 
Plugin it into expression for $m^2_{eff}$ and keeping terms up to the first order in 
$\epsilon$ and $\delta_H$ we obtain
\begin{equation}
m^2_{eff} = - \frac{1}{\tau^2}\left[ 2+3\epsilon\left(1- 3\delta_H\right) \right]. 
\end{equation}
The equation for tensor modes can be therefore expressed as
\begin{equation}
\frac{d^2 u_k}{d\tau^2} +\left[  \Omega k^2 -\left(\nu_T^2 -\frac{1}{4} \right) \frac{1}{\tau^2} \right] u_k =0, 
\label{EOMTensorHankel}
\end{equation}
where 
\begin{equation}
|\nu_T| = \sqrt{ \frac{9}{4}+3\epsilon \left(1-3 \delta_H  \right)} 
\simeq \frac{3}{2}+\epsilon  \left(1-3 \delta_H \right).
\end{equation}

Equation (\ref{EOMTensorHankel}) reminds the standard equation for inflationary modes and 
it is tempting to find its analytic solution in terms of Hankel functions.  This however would not 
be consistent because requires assumption of constancy of $\Omega$. The slow variation of 
$\Omega$ cannot be neglected if we already included variation of $\Omega$ in the expression 
for $m^2_{eff}$. To see it clearly, let us perform the following change of variables:
\begin{eqnarray}
x &:=& -k\tau \sqrt{\Omega}, \\ 
f(x) &:=& \frac{u}{\sqrt{x}}, 
\end{eqnarray}
which transforms (\ref{EOMTensorHankel}) into 
\begin{equation}
(1-4\epsilon \delta_H) x^2\frac{d^2f}{dx^2}+x\frac{df}{dx}+(x^2-\nu_T^2)f = 0. 
\label{EquationModesF} 
\end{equation}
While the dependence on $\delta_H$ in $m^2_{eff}$ is contributing to $\nu_T^2$, the 
time dependence of $\Omega$ in front of $k^2$ generates factor $-4\epsilon \delta_H$. 
Because of this factor, solutions to equation (\ref{EquationModesF}) are not Bessel (or equivalently 
Hankel) functions, what would be the case if the factor $-4\epsilon \delta_H$ is absent. 
Nevertheless, solutions to equation Ref. (\ref{EOMTensorHankel}) can be studied numerically. 

Because analytic solution to equation (\ref{EOMTensorHankel}) cannot be easily found, 
we are forced to use another approach to find tensor power spectrum. Namely we will 
determine amplitude of the perturbations at the Hubble radius with use of the short scale 
solutions. However first, in order to approve consistency of normalization in case on 
non-vanishing $\Omega$ we will consider tensor power spectrum for the case with 
$\Omega=$ const. 
 
\subsection{$\Omega=$ const case} 

As far as $\Omega$ can be considered as a constant, the effective mass term is
\begin{equation}
m^2_{eff} \equiv -\frac{z_T ^{''}}{z_T}=- \frac{a^{''}}{a}= -\frac{1}{\tau^2}\left[2+3\epsilon\left(1- \delta_H\right)\right].
\end{equation}
The equation of motion takes the form (\ref{EOMTensorHankel}) with $\Omega=$const and 
\begin{equation}
|\nu_T| = \sqrt{ \frac{9}{4}+3\epsilon \left(1-\delta_H  \right)} 
\simeq \frac{3}{2}+\epsilon  \left(1-\delta_H \right).
\end{equation}
In this case, exact solution to equation (\ref{EOMTensorHankel}) can be expressed in terms of Hankel functions: 
\begin{equation}
u_k = \sqrt{-\tau} \sqrt{\frac{\pi}{4}} \left[  D_1 H^{(1)}_{|\nu|}(-\sqrt{\Omega} k \tau)+
D_2 H^{(2)}_{|\nu|}(-\sqrt{\Omega} k \tau) \right]. 
\end{equation}
The constants $D_1$ and $D_2$ were normalized such chosen such that the Wronskian condition (\ref{Wronskian}) 
leads to relation
\begin{equation}
|D_1|^2-|D_2|^2=1. 
\end{equation}
The $\Omega-$deformed Minkowski vacuum normalization is chosen by taking  $D_2 = 0$ and 
$D_1 =e^{i\pi(2|\nu|+1)/4}$.  This can be verified by considering asymptotic behavior of the Hankel 
function. Namely, for $x\ll 1$ $H^{(1)}_{|\nu|}(x) \approx   \sqrt{\frac{2}{\pi x}} \exp \left( i(x-|\nu|\pi/2-\pi/4) \right) $. 
With use of this   
\begin{equation}
u_k = \sqrt{-\tau} \sqrt{\frac{\pi}{4}} e^{i\pi(2|\nu|+1)/4} H^{(1)}_{|\nu|}(-\sqrt{\Omega} k \tau) \approx
\frac{e^{-i \sqrt{\Omega}k \tau }}{\sqrt{2\sqrt{\Omega}k}} = u_k^{\Omega M},
\end{equation}
for $-\sqrt{\Omega} k \tau \gg 1$.  

Having the modes correctly normalized we can study the super-horizonal limit $-\sqrt{\Omega} k \tau \ll 1$.   
With use of approximation  $H^{(1)}_{|\nu|}(x) \simeq - \frac{i}{\pi} \Gamma(|\nu|) \left(\frac{x}{2} \right)^{-|\nu|}$,
which holds at $x\ll1$, we obtain 
\begin{equation}
|u_k|^2 \simeq  \frac{1}{2} \frac{1}{aH} \left( \frac{k \sqrt{\Omega}}{aH} \right)^{-2|\nu|}, 
\end{equation}
where we used $-\tau \simeq \frac{1}{aH}$. Applying it to definition (\ref{TensorPower}), the tensor power spectrum from 
the slow-roll inflation is:
\begin{equation}
\mathcal{P}_{T}(k) = A_T \left( \frac{k}{aH}\right)^{n_{T}},
\end{equation}   
where the amplitude 
\begin{equation}
A_T = \frac{16}{\pi} \left(\frac{H}{m_{\text{Pl}}} \right)^2\left(1+\delta_H \right),
\label{Attheo}
\end{equation}
and the tensor spectral index
\begin{eqnarray}
n_{T} =-2\epsilon \left(1-\delta_H\right).
\end{eqnarray}

With use of the modified Friedmann equation (\ref{Friedmann}) in the slow-roll regime 
$(\rho \approx V)$, one can rewrite (\ref{Attheo}) into the following form
\begin{equation}
A_T = \frac{128}{3}  \frac{V}{ \rho_{Pl}}\left(1- \delta_H\right) \left(1+\delta_H   \right)
= \frac{128}{3} \frac{V}{ \rho_{Pl}}+\mathcal{O}(\delta_H^2). 
\end{equation}
This expression is not a subject of holonomy corrections in the leading order.

\subsection{$\Omega-$deformed Minkowski vacuum} 

Let us now proceed to the proper calculations. The strategy is the following: We will use 
a given short scale solution and extrapolate it up to the horizon scale. A mode characterized by
$k$ crosses the horizon scale when $\sqrt{\Omega} k = aH \simeq -\frac{1}{\tau}$. 
Above the horizon scale the modes ``freeze out'' and the power spectrum remains unchanged. 
The spectral index can be computed from the horizontal spectrum based on the formula 
\begin{equation}
n_{T} \equiv \frac{d \ln \mathcal{P}_T}{d \ln k}.  
\label{nTdeff}
\end{equation}

Modulus square of the  $\Omega-$deformed  Minkowski vacuum (\ref{WKBHoloMink}) is
\begin{equation}
\left|v^{\Omega M}_k \right|^2  =\frac{1}{2\sqrt{\Omega}k}. 
\label{ModSqvOM}
\end{equation}
By inserting (\ref{ModSqvOM}) into the definition (\ref{TensorPower}) and calculating 
the value at $k \sqrt{\Omega} =a H$ we find 
\begin{equation}
\mathcal{P}_{T}(k) =  \frac{16}{\pi} \left(\frac{H}{m_{Pl}} \right)^2\frac{1}{\sqrt{\Omega}} 
= \frac{16}{\pi} \left(\frac{H}{m_{Pl}} \right)^2(1+\delta_H)+\mathcal{O}(\delta_H^2), 
\end{equation}   
which agrees with (\ref{Attheo}). 

Let us this result and  to compute tensor spectral index. By using (\ref{nTdeff}), we find
\begin{equation}
n_{T} = 2 \frac{k}{H} \frac{dH}{dk}- \frac{1}{2}\frac{k}{\Omega}\frac{d\Omega}{dk}
= - 2\epsilon +\mathcal{O}(\epsilon^2\delta_H),   
\label{nTOM}
\end{equation}
where we used $k = -\frac{1}{\tau \sqrt{\Omega}}$, (\ref{SRepsilon}) and  (\ref{OmegaPrimeOmega}).
Here, the correction from $\delta_H$ contributes together with $\epsilon^2$ terms. Therefore, in the 
leading order the tensor spectral index holds its classical form. 
  
\subsection{Bojowald-Calcagni normalization}

Let us now compute the power spectrum for the Bojowald-Calcagni normalization. 
While use of condition $\sqrt{\Omega} k = aH \simeq -\frac{1}{\tau}$, the modulus square 
of $v^{BC}_k$ gives  
\begin{equation}
\left| v^{BC}_k \right|^2 \simeq \frac{1}{2k}(1+2\delta_H). 
\label{ModSqvBC}
\end{equation}
It is wort mentioning that this value does not depend on the fact that the 
amplitude is computed at the horizon. This is because the $i k \tau \delta_{H}$ term 
contributes in the second order, which is neglected. By inserting (\ref{ModSqvBC})
into the definition (\ref{TensorPower}) we find 
\begin{equation}
\mathcal{P}_{T}(k) = \frac{16 G}{\pi}  \frac{k^2}{a^2} \Omega (1+2\delta_H)
= \frac{16}{\pi} \left(\frac{H}{m_{Pl}} \right)^2(1+2\delta_H)+\mathcal{O}(\delta_H^2). 
\end{equation}   

Having amplitude of spectrum computed at the horizon scale, the spectral index is computed 
from the relation
\begin{eqnarray}
n_{T} \equiv \frac{d \ln \mathcal{P}_T}{d \ln k} = 2 \frac{kHa}{ dk/d\tau} \frac{\dot{H}}{H^2}+ 2\frac{d\delta_H}{d\ln k} 
= -2 \epsilon (1+\delta_H),
\end{eqnarray}
where we have used expression  (\ref{SRepsilon}) and the fact that at the horizon $\sqrt{\Omega} k\tau=-1$.

Summing up, the tensor power spectrum with the Bojowald-Calcagni normalization can be 
expressed as follows 
\begin{equation}
\mathcal{P}_{T}(k) = A_T \left( \frac{k}{aH}\right)^{n_{\text{T}}},
\end{equation}   
where the amplitude 
\begin{equation}
A_T = \frac{16}{\pi} \left(\frac{H}{m_{Pl}} \right)^2(1+2\delta_H)
\label{AttheoBC}
\end{equation}
and the tensor spectral index
\begin{eqnarray}
n_{T} =-2\epsilon  \left(1+\delta_H\right).
\label{nTBC}
\end{eqnarray}

With use of the modified Friedmann equation (\ref{Friedmann}) in the slow-roll regime 
$(\rho \approx V)$, one can rewrite (\ref{AttheoBC}) into the following form
\begin{equation}
A_T = \frac{128}{3}  \frac{V}{ \rho_{Pl}}   \left(1+\delta_H\right). 
\end{equation}

\section{Scalar power spectrum}

Here, for the sake of completeness we will derive equation of motion for the scalar modes in the
slow-roll approximation. This equation will not be used to derive spectrum of the scalar 
inflationary perturbations because of the same reason as in the case of tensor modes.  

Amplitude of the scalar power spectrum 
\begin{equation}
\mathcal{P}_S(k) = \frac{k^3}{2\pi^2} \left| \frac{v_k}{z_S} \right|^2
\label{ScalarPower}
\end{equation}
will be calculated using the short scale solution extrapolated to the horizontal scale. While the spectral 
scalar is found, the spectral index will be determined by virtue of 
\begin{equation}
n_{S} \equiv \frac{d \ln \mathcal{P}_S}{d \ln k}.  
\label{nSdeff}
\end{equation}

Similarly as for tensor modes, the Fourier transform of the scalar perturbations fulfills equation 
\begin{equation}
\frac{d}{d\tau^2}v_k+\Omega k^2 v_k-\frac{z^{''}_S}{z_S}v_k=0, 
\end{equation}
where $z_S = a  \frac{ \varphi'}{\mathcal{H}}$. The task now is to determine time dependance of 
$\frac{z^{''}_S}{z_S}$ in the slow-roll approximation.  

By differentiating $z_S = a  \frac{ \varphi'}{\mathcal{H}}$ with respect to conformal time and 
by using relation  (\ref{SRepsilon}) we obtain
\begin{equation}
\frac{z'_S}{z_S} = \epsilon \left(1-\delta_H\right) \mathcal{H}+\frac{\varphi^{''} }{\varphi'}. 
\label{zbSzS}
\end{equation}
With use this, the expression (\ref{deltaSR}) for the parameter $\delta$ can be written as
\begin{equation}
\delta = 1- \frac{\varphi^{''} }{\varphi'\mathcal{H}} = 1+\epsilon \left(1-\delta_H\right) -\frac{z'}{z\mathcal{H}}.  
\label{deltaSR2}
\end{equation}
By differentiating this equality with respect to conformal  time and neglecting all the non-leading contributions 
(\emph{i.e.} $\delta', \epsilon', \epsilon^2, \eta'$) we obtain the following equality
\begin{equation}
\frac{z^{''}_S}{z_S} = \left(\frac{z^{'}_S}{z_S} \right)^2 +\frac{\mathcal{H}'}{\mathcal{H}}  \frac{z^{'}_S}{z_S}. 
\label{zSbbzS1}
\end{equation}
Combining (\ref{zbSzS}) together with (\ref{deltaSR2}) and (\ref{deltaSR}) we find 
\begin{equation}
\frac{z^{'}_S}{z_S}  = \left[1-\eta+2\epsilon(1-\delta_H)\right]\mathcal{H}.
\label{zbSzS2} 
\end{equation}
Furthermore 
\begin{equation}
\frac{\mathcal{H}'}{\mathcal{H}} = \mathcal{H}\left(1+\frac{\dot{H}}{H^2}\right)
= \mathcal{H}\left(1-\epsilon(1-\delta_H)\right),
\label{dHH}
\end{equation} 
where in the second equality we used (\ref{SRepsilon}). Plugging (\ref{zbSzS2}) and (\ref{dHH}) 
to (\ref{zSbbzS1}) we obtain
\begin{eqnarray}
\frac{z^{''}_S}{z_S} &=& \mathcal{H}^2\left[2+5\epsilon(1-\delta_H)-3\eta \right]
= \frac{1}{\tau^2\left[1-\epsilon(1-\delta_H) \right]^2}\left[2+5\epsilon(1-\delta_H)-3\eta  \right]      \nonumber\\ 
&=& \frac{1}{\tau^2} \left[2+9\epsilon(1-\delta_H)-3\eta\right]. 
\label{zSbbzS2}
\end{eqnarray}

Equation for scalar modes with holonomy correction in the (first order) slow-roll approximation
can be therefore written as 
\begin{equation}
\frac{d^2 v_k}{d\tau^2} +\left[  \Omega k^2 -\left(\nu_S^2 -\frac{1}{4} \right) \frac{1}{\tau^2} \right] v_k =0, 
\label{EOMScalarHankel}
\end{equation}
where 
\begin{equation}
|\nu_S| = \sqrt{ \frac{9}{4}+9\epsilon \left(1-3 \delta_H  \right)-3\eta} 
\simeq \frac{3}{2}+3\epsilon  \left(1-3 \delta_H \right)-\eta.
\end{equation}
The classical case is correctly recovered for $\delta_H \rightarrow 0$.

\subsection{$\Omega-$deformed Minkowski vacuum} 

With use of the WKB approximation (\ref{WKBHoloMink}) applied to definition (\ref{ScalarPower})
we find 
\begin{equation}
\mathcal{P}_{S}(k) = 
\frac{1}{\pi \epsilon} \left(\frac{H}{m_{Pl}} \right)^2  \frac{(1-\delta_H)}{\Omega^{3/2}}=
\frac{1}{\pi \epsilon} \left(\frac{H}{m_{Pl}} \right)^2 \left(1+2 \delta_H  \right)+\mathcal{O}(\delta_H^2)
\end{equation}
at the horizon $\sqrt{\Omega} k = aH$. 

Finally, the inflationary scalar power spectrum:
\begin{equation}
\mathcal{P}_{S}(k) = A_S\left( \frac{k}{aH}\right)^{n_{S}-1}, 
\end{equation}
where amplitude of the scalar perturbations
\begin{equation}
A_S = \frac{1}{\pi \epsilon} \left(\frac{H}{m_{Pl}} \right)^2 \left(1+2 \delta_H  \right),
\label{Astheo}
\end{equation}
and the spectral power index
\begin{equation}
n_{S} =1+2\eta-6\epsilon +\mathcal{O}(\epsilon^2\delta_H,\epsilon\eta\delta_H). 
\end{equation}

As in case of the tensor modes, the $\delta_H$ correction to the spectral 
index is multiplied by the $\epsilon^2$ and $\epsilon\eta$ factors 
which are negligible in the considered order. To see it explicitly let us consider 
the case of massive scalar field ($V=\frac{1}{2}m^2\varphi^2$) for which $\epsilon=\eta$. 
Therefore $n_{S} =1-4\epsilon+\mathcal{O}(\epsilon^2\delta_H)$. Using the 
recent Planck fit $n_S= 0.9603\pm0.0073$ \cite{Ade:2013uln}, we obtain 
$\epsilon \approx \frac{1}{4}(1-n_S) \approx  0.01$. The higher order corrections are 
therefor of the order $\mathcal{O}(\epsilon^2\delta_H,\epsilon\eta\delta_H)\sim 10^{-4}\delta_H$, 
with $|\delta_H| < \frac{1}{2}$. These terms are also typically smaller than 
contributions from the classical second order slow-roll expansion. With use of 
the present observational precision is impossible to constrain such effects.    

Moreover, with use of the modified Friedmann equation (\ref{Friedmann}) in the slow-roll 
regime $(\rho \approx V)$, one can rewrite (\ref{Astheo}) into the following form
\begin{equation}
A_S =  \frac{8}{3} \frac{1}{\epsilon}   \frac{V}{ \rho_{Pl}}   \left(1-\delta_H \right) \left(1+2 \delta_H  \right)
\approx \frac{8}{3} \frac{1}{\epsilon}   \frac{V}{ \rho_{Pl}} \left(1+\delta_H \right).
\end{equation}

\subsection{Bojowald-Calcagni normalization}

The calculations can be now repeated for the case of Bojowald-Calcagni normalization. 
The obtained inflationary scalar power spectrum is 
\begin{equation}
\mathcal{P}_{S}(k) = A_S\left( \frac{k}{aH}\right)^{n_{S}-1},
\end{equation}
where amplitude of the scalar perturbations   
\begin{equation}
A_S = \frac{1}{\pi \epsilon} \left(\frac{H}{m_{Pl}} \right)^2     \left(1+3\delta_H  \right)
\label{Astheo}
\end{equation}
and the spectral index
\begin{equation}
n_{S} =1+2\eta-6\epsilon \left(1+\frac{1}{3}\delta_H \right). 
\end{equation}
In contrary to the previous case, the spectral index is holonomy-corrected 
in the leading order for the Bojowald-Calcagni normalization

For completeness, with use of the modified Friedmann equation (\ref{Friedmann}) in the slow-roll regime 
$(\rho \approx V)$, one can rewrite (\ref{Astheo}) into the following form
\begin{equation}
A_S =  \frac{8}{3} \frac{1}{\epsilon}   \frac{V}{ \rho_{Pl}} \left(1-\delta_H\right) \left(1+3 \delta_H  \right)
\approx \frac{8}{3} \frac{1}{\epsilon}   \frac{V}{ \rho_{Pl}} \left(1+2\delta_H \right).
\end{equation}

\section{Tensor-to-scalar ratio}

In theoretical studies of inflation as well in confronting theoretical predictions with observations it is 
often useful to work with tensor-to-scalar ratio $r$. This dimensionless quantity, defined as  
\begin{equation}
r :=\frac{A_T}{A_S},
\end{equation} 
measures ratio between amplitudes of tensor and scalar perturbations. There is at present a huge effort 
to detect B-type polarization of the CMB radiation which would make determination of the amplitude 
of the tensor perturbations $A_T$ possible\footnote{To be precise, the B-type polarization of the 
primordial origin was not detected yet. The B-type polarization due to gravitational lensing was recently 
observed for the first time by the SPTpol observatory \cite{Hanson:2013hsb}.}. At present, knowing 
the value of $A_S$ and having observational constraint on $A_T$, upper bound on the value of $r$ 
can be found. The strongest constraint comes from observations of the Planck 
satellite:  $r < 0.11$ (95\% CL) \cite{Ade:2013uln}. The theoretically predicted values of $r$ can be 
confronted with this bound allowing for elimination of some possible inflationary scenarios. In particular,
the massive model of inflation is no more preferred in the light of the new Planck constraint \cite{Ade:2013uln,Ijjas:2013vea}.     

Let us calculate the tensor-to-scalar ratio $r$ for the models studied in this paper. For the case with 
$\Omega-$deformed Minkowski vacuum normalization we obtain 
\begin{equation}
r = 16\epsilon \frac{\left(1+ \delta_H \right)}{\left(1+2 \delta_H \right)} = 
16\epsilon\left(1- \delta_H \right) +\mathcal{O}(\delta_H^2).
\label{rOM}
\end{equation} 
Based on this and  equation (\ref{nTOM}) the expression for the tensor spectral index
\begin{equation}
r \approx - 8 n_{T}  \left(1-\delta_H \right).
\end{equation}

As we have shown in the previous section, for the massive scalar field $\epsilon \approx 0.01$.
For the classical case ($\delta_H=0$) this would give us $r = 16\epsilon  = 0.16$, which is in contradiction with 
the Planck constraint $r < 0.11$. This reflects the mentioned disagreement between the massive 
scalar field model of inflation and the new Planck data. In the past, when the observational bound 
on the value of $r$ was weaker, the massive scalar field model of inflation was favored by the data.  
It is worth noticing that, by applying $\epsilon \approx 0.01$ to (\ref{rOM}), together with the Planck 
constraint on $r$, we find that $\delta_H \gtrsim 0.3$.  Therefore, presence of the quantum holonomy 
corrections helps to fulfill the observational bound. However, this would require the critical energy density 
$\rho_{\text{c}}$ to be much smaller than the Planck energy density.   

For the Bojowald-Calcagni normalization we obtain 
\begin{equation}
r = 16\epsilon \frac{\left(1+2\delta_H \right)}{\left(1+3 \delta_H \right)} = 16\epsilon\left(1- \delta_H \right) +\mathcal{O}(\delta_H^2),
\end{equation} 
which is the same as for the $\Omega-$deformed Minkowski vacuum normalization. Finally, based on this and 
(\ref{nTBC}) the expression for the tensor spectral index
\begin{equation}
r \approx - 8 n_{T}  \left(1-2\delta_H \right).
\end{equation}

\section{Summary}

In this paper we found holonomy corrections to inflationary power spectra. 
Such corrections reflect a discrete nature of space at the Planck scale predicted 
by loop quantum gravity. Calculations were performed for the slow-roll type inflation 
driven by a single self-interacting scalar field.  The derivations were done up to the 
first order in the slow-roll parameters $\epsilon$ and $\eta$ as well as in the leading 
order in the parameter $\delta_H$, characterizing holonomy corrections.

An important issue while considering quantum fields on expanding backgrounds 
is a proper normalization of the modes. In our calculations we assumed that only ingoing  
modes are present. Short scale normalization of these modes is a subject of ambiguity 
due to presence of the quantum holonomy effects. We considered two, best 
motivated, types of normalization.  The first one was based on adiabatic vacuum 
(WKB) approximation, while the second one was based on the method proposed 
by Bojowald and Calcagni in Ref. \cite{Bojowald:2010me}.  

For the first type of normalization, spectral indices are not quantum corrected in the 
leading order. To be precise, linear corrections in $\delta_H$ are expected. However,
they are multiplied not by $\epsilon$ or $\eta$ but $\epsilon^2$ or $\eta^2$ terms. These 
higher order contributions were not studied systematically in this paper. Nevertheless,
calculation of the holonomy-corrected inflationary spectrum including 
$\mathcal{O}(\epsilon^2, \eta\epsilon,\eta^2)$ terms is a natural generalization of the 
results presented here. This would help constraining $\delta_H$ if sufficiently accurate 
observational data are available. Investigation of the higher order corrections 
in the light of the present CMB data is, however, not possible.      

As we have shown, equation of motion for tensor modes with holonomy corrections 
can be derived from the wave equation defined on \emph{effective metric}, which encodes 
quantum gravitational effects. This observation allowed us to define a proper inner product 
for the tensor modes and to show that normalization of tensor modes is obtained by satisfying 
the classical Wronskian condition.  
 
In this paper we focused on the region where $\Omega>0$. Much more interesting is
behavior of modes in the vicinity of $\Omega=0$ and for  $\Omega<0$ where the 
equations of modes become elliptic. The issue of imposing initial conditions at $\Omega=0$ 
will be a subject of the forthcoming paper \cite{MielczarekNEW}. Evolution of tensor 
modes across the region with negative $\Omega$ was addressed in Ref. \cite{Linsefors:2012et}. 
As it was shown there, tensor power spectrum is enormously amplified in the UV regime. 
This new behavior certainly deserves more detailed studies.  Furthermore, investigation of 
simultaneous effects of holonomy and inverse volume corrections is now possible
thanks to new results presented in Ref. \cite{Cailleteau:2013kqa}.   

\acknowledgments
I would like to thank Gianluca Calcagni for useful discussion.

\end{document}